\documentclass[aps,prd,reprint,superscriptaddress,amsmath,amssymb,nofootinbib,floatfix]{revtex4-2}
\usepackage{graphicx}
\usepackage{dcolumn}
\usepackage[utf8]{inputenc}
\usepackage{multirow}

\usepackage{ulem}
\usepackage{xcolor}
\usepackage[unicode=true,colorlinks,linkcolor=blue,anchorcolor=blue,urlcolor=blue,citecolor=blue,breaklinks=true]{hyperref}
\usepackage{orcidlink}
\usepackage{subfigure}

\newcommand{\ii}{i} 
\newcommand{\ee}{e} 
\newcommand{\pararrow}{\mathord{\buildrel{\lower3pt\hbox{$\scriptscriptstyle\leftrightarrow$}}\over {\partial}}} 
\newcommand{\pararrowk}[1]{\mathord{\buildrel{\lower3pt\hbox{$\scriptscriptstyle\leftrightarrow$}}\over {\partial}\hspace*{-0.18em}{}^#1}\hspace*{-0.18em} \,} 
\newcommand{\olsi}[1]{\,\overline{\!{#1}}}

\usepackage{physics}
\usepackage{bm}
\usepackage{easyReview}

\newcommand{\qfnu}{\affiliation{College of Physics and Engineering, Qufu Normal University, Qufu 273165, China}}
\newcommand{\IFIC}{\affiliation{Instituto de Física Corpuscular (centro mixto CSIC-UV),\\ Institutos de Investigación de Paterna, Apartado 22085, 46071, Valencia, Spain}}

\begin{document}

\title{ Radiative decays of the isoscalar $S$-wave $D\olsi D$ molecule }
    
    \author{Yuan-Jun Gao\,\orcidlink{0009-0009-2616-4816}} \qfnu
	\author{Gang Li\,\orcidlink{0000-0002-5227-8296}} \email{gli@qfnu.edu.cn} \qfnu
	\author{Shi-Dong Liu\,\orcidlink{0000-0001-9404-5418}} \email{liusd@qfnu.edu.cn} \qfnu
    \author{Pan-Pan Shi\,\orcidlink{0000-0003-2057-9884}} 
    \email{Panpan.Shi@ific.uv.es} \IFIC
 
\begin{abstract}
In the hadronic molecular picture, an isoscalar $S$-wave $D\olsi{D}$ molecule is naturally expected as the spin-0 partner of $X(3872)$. Assuming this state, denoted by $X_0$, to be a pure $D\olsi D$ molecule with quantum numbers $J^{PC}=0^{++}$, we investigate its radiative decays $X_0\to\gamma V (V=\rho^0,\omega)$ within an effective Lagrangian framework. The decay amplitudes are generated by intermediate charmed-meson loops, with electromagnetic gauge invariance consistently maintained throughout the calculation. We evaluate the partial decay widths and examine their dependence on the $X_0$ mass and the cutoff. Although the individual decay widths exhibit a sizable dependence on the cutoff, their ratio is remarkably stable. In particular, we obtain that the ratio for the decays $X_0\to\gamma\rho^0$ and $X_0\to\gamma\omega$ is approximately 2.26, which is insensitive to the variation of the cutoff. This robust ratio provides a useful model-insensitive signature of the molecular structure of $X_0$.
\end{abstract}

\date{\today}

\maketitle

\section{Introduction} \label{sec:intro}

The conventional quark model, built upon the fundamental $q\olsi q$ meson and $qqq$ baryon configurations, provides a successful description of the bulk of the hadron spectrum and serves as a theoretical framework for understanding the structures of hadronic states in the nonperturbative regime of quantum chromodynamics (QCD). In the hidden-charm sector, however, the discovery of the $X(3872)$ by the Belle Collaboration in 2003~\cite{Belle:2003nnu} marked the beginning of extensive investigations of the so-called $XYZ$ states, for recent reviews, see Refs.~\cite{Hosaka:2016pey,Esposito:2016noz,Lebed:2016hpi,Ali:2017jda,Olsen:2017bmm,Guo:2017jvc,Albuquerque:2018jkn,Liu:2019zoy,Guo:2019twa,Brambilla:2019esw,Chen:2022asf,Liu:2024uxn,Chen:2024eaq,Wang:2025sic,Dai:2026fkg}.
The experimental measurements of this state posed a substantial challenge to the conventional charmonium interpretation.
Although the quantum numbers of $X(3872)$ are consistent with a conventional $2P$ charmonium, its mass is extremely close to the $D^0\olsi D{}^{*0}$ threshold, which is far below quark-model predictions~\cite{Eichten:1979ms,Godfrey:1985xj,Barnes:2005pb}. Furthermore, the ratio of the decay rates of $X(3872)$ into the isospin-violating $J/\psi \rho^0$ and isospin-conserving $J/\psi \omega$ channels provides a sensitive probe of isospin breaking in the X(3872). The experimentally observed ratio is unexpectedly large compared with the expectation for a pure charmonium state~\cite{LHCb:2022jez}, indicating substantial isospin violation in its decay dynamics and disfavoring a purely charmonium interpretation. These features strongly motivate a hadronic molecule interpretation of the $X(3872)$, with its proximity to the $D^0 \olsi D{}^{*0}$ threshold favoring a $D \olsi D{}^{*}$ molecular component and naturally enhancing isospin-breaking effects. 

A crucial test of the hadronic molecular interpretation of the $X(3872)$ is the search for its spin partners. One such candidate is the isoscalar $D\olsi D$ bound state with quantum numbers $J^{PC}=0^{++}$, denoted as $X_0$ hereafter, whose existence is motivated by heavy-quark spin symmetry (HQSS)~\cite{Nieves:2012tt,Hidalgo-Duque:2012rqv,Guo:2013sya,Baru:2017pvh}. The possibility of an isoscalar $D\olsi D$ bound state was first explored within a vector-meson-exchange potential~\cite{Zhang:2006ix} and was subsequently studied using the coupled-channel unitary approaches~\cite{Gamermann:2006nm}. The $X_0$ was also studied in various phenomenological frameworks~\cite{Liu:2009qhy,Liu:2017mrh,Dong:2021juy,Li:2022shq,Abreu:2025jqy,Ji:2022uie,Ji:2022vdj}.

Several lattice studies have investigated the possible structures in the $D\olsi D$ system. Using the ensembles with pion mass $266$ and $156~\mathrm{MeV}$, Ref.~\cite{Lang:2015sba} found evidence for a virtual state located about $20~\mathrm{MeV}$ below the $D\olsi D$ threshold.\footnote{A discussion of this state was subsequently provided in Ref.~\cite{Prelovsek:2020eiw}.} The subsequent study of the $D\olsi D$--$D_s\olsi D{}_s$ coupled-channel system with pion mass about $280~\mathrm{MeV}$ found three $0^{++}$ states. One of these states, below the $D\olsi D$ threshold by about $4~\mathrm{MeV}$, was identified as an isoscalar $D\olsi D$ bound state. A bound state was also obtained in the reanalysis of the above lattice data within the framework of the effective field theory~\cite{Shi:2024llv}. In contrast, lattice studies with pion mass $391~\mathrm{MeV}$ found no bound or resonant state close to the $D\olsi D$ threshold~\cite{Wilson:2023hzu,Wilson:2023anv}, although a distant virtual state was mentioned in some parametrizations~\cite{Wilson:2023anv}. More recently, the same authors studied the $\eta\eta_c$--$D\olsi D$ coupled-channel system at pion masses of $239$, $283$, and $330~\mathrm{MeV}$ and found no bound state or resonance near the $D\olsi D$ threshold~\cite{Wilson:2026bhu}. These lattice results therefore do not yet provide a consistent picture of a near-threshold $X_0$, highlighting the need for further investigation of the $D\olsi D$ system.

Experimental efforts have been devoted to the search for the $X_0$. In the $e^+e^-\to J/\psi\,D\olsi D$ process, Belle observed a bump near the $D\olsi D$ threshold~\cite{Belle:2007woe}, which was subsequently interpreted as a possible $D\olsi D$ bound state~\cite{Gamermann:2007mu,Wang:2019evy}. A near-threshold enhancement was also observed in $\gamma\gamma\to D\olsi D$ by Belle and BABAR~\cite{Belle:2005rte,BaBar:2010jfn} and proposed as a possible $D\olsi D$ bound state.  However, these observed enhancements can also be interpreted as the contributions of the conventional charmonium spectrum. In particular, the enhancement in the $\gamma\gamma\to D\olsi D$ channel could be attributed to the $\chi_{c0}(2P)$~\cite{Guo:2012tv}; the bump in the $e^+e^-\to J/\psi\,D\olsi D$ channel was suggested to originate from the $\chi_{c0}(2P)$ followed by the observation of the $\chi_{c0}(3860)$~\cite{Belle:2017egg}. More recently, a combined reanalysis of the $\gamma\gamma\to D\olsi D$~\cite{Belle:2005rte,BaBar:2010jfn}, $B^{+}\to K^+ D^+ D^-$~\cite{LHCb:2020pxc}, and $B^+\to K^+ D_s^+ D_s^-$~\cite{LHCb:2022aki} data reported evidence for the $X_0$~\cite{Ji:2022vdj}. On the other hand, BESIII searched for the $X_0$ in $\psi(3770)\to\gamma\eta\eta'$ but found no significant structure in the $\eta\eta'$ invariant-mass spectrum~\cite{BESIII:2023bgk}. These results leave the experimental status of the $X_0$ unsettled and motivate further searches in complementary decay channels and production processes.

To facilitate the search for the $X_0$ (also referred to as $X(3700)$ or $X(3720)$ in some references), various production and decay channels have been investigated. Radiative decays of the $\psi(3770)$ have been proposed as promising production mechanisms for the $X_0$~\cite{Gamermann:2009ouq,Dai:2020yfu}. Near-threshold enhancement associated with the $X_0$ have been proposed in the $D\olsi D$ invariant-mass distributions from various reactions, including $e^+e^-\to J/\psi D\olsi D$~\cite{Gamermann:2007mu,Wang:2019evy}, $B\to D\olsi DK$~\cite{Dai:2015bcc,Wang:2026mfi,Ren:2026kai}, $\gamma\gamma\to D\olsi D$~\cite{Wang:2020elp,Deineka:2021aeu,Sobrinho:2026agd}, and $\Lambda_b\to\Lambda D\olsi D$~\cite{Wei:2021usz,Ren:2026kai}. The $X_0$ was also suggested to be reconstructed through the $\eta\eta$ invariant mass distribution in the process $B^+\to K^+\eta\eta$~\cite{Brandao:2023vyg}, through the $\eta\eta^{\prime}$ invariant mass distribution in the process $\psi(3770)/\psi(4040)\to\gamma\eta\eta^\prime$ and $e^+e^-\to J/\psi\eta\eta^\prime$~\cite{Xiao:2012iq}, and through the $\eta\eta_c$ invariant mass distribution in the process $B^-\to K^-\eta\eta_c$~\cite{Li:2023nsw}. In addition, the radiative decay $X_0\to\gamma J/\psi$~\cite{Gamermann:2007bm}, and its hadronic decays into pairs of light pseudoscalar and vector mesons~\cite{Gao:2025zhp} have also been discussed.

Despite these studies, several potentially informative decay channels of the $X_0$ remain unexplored. Radiative decays are particularly valuable for probing the internal structure of $D^{(*)}\olsi D{}^{(*)}$ systems, as demonstrated in studies of the $X(3872)$ and its spin-2 partner $X_2$~\cite{Dong:2009uf,Guo:2014taa,Albaladejo:2015dsa,Shi:2023mer,Shi:2023ntq,Wang:2025zss}. In particular, radiative decay ratios of the $X(3872)$ have been shown to be sensitive to the admixture of molecular and $c\olsi c$ components~\cite{Dong:2009uf,Guo:2014taa}, while the radiative decay $X_2\to D\olsi D{}^*\gamma$ is sensitive to the long-range structure of the state and can therefore help distinguish different internal configurations~\cite{Albaladejo:2015dsa}. Other studies have identified characteristic radiative observables that may be useful in searches for the $X(3872)$ and $X_2$~\cite{Shi:2023mer,Wang:2025zss}. Motivated by these results, we investigate the radiative decays $X_0\to\gamma V$, with $V=\rho^0,\omega$, through intermediate charmed-meson loops. The resulting predictions provide theoretical benchmarks for future experimental searches for the $X_0$ and may offer additional insight into its internal structure.

The remainder of this paper is organized as follows. We introduce the effective Lagrangians and transition amplitudes for the radiative decay $X_0\to\gamma V$ in Sec.~\ref{sec:formula}. Numerical results and relevant discussions are presented in Sec.~\ref{sec:result}. Finally, a summary is given in Sec.~\ref{sec:summary}.

\section{Formalism} \label{sec:formula}

\subsection{Effective Lagrangian for the relevant interactions}

In the pure hadronic molecular scenario, $X_0$ is treated as an isoscalar $S$-wave $D\olsi D$ bound state. 
The interaction between $X_0$ and $D \olsi D$ pair is described as
\begin{equation} \label{eq:LagX0}
    \mathcal{L}_{X_0} = X_0^\dagger (\chi^0_{\mathrm{nr}} D^{0} \olsi{D}{}^{0} + \chi^c_{\mathrm{nr}} D^{+} D^{-}) + \mathrm{H.c.}\,,
\end{equation}
where $\chi^0_{\mathrm{nr}}$ and $\chi^c_{\mathrm{nr}}$ denote the coupling strengths of $X_0$ to the neutral and charged charmed-meson channels, respectively. In this work, we set $\chi^0_{\mathrm{nr}}=\chi^c_{\mathrm{nr}}=\chi_{\mathrm{nr}}$. For a shallow molecular state, the effective coupling constant is related to its binding energy~\cite{Weinberg:1965zz, Baru:2003qq}
\begin{equation} \label{eq:coupleX0}
	\chi_{\mathrm{nr}} = \left(\frac{16 \pi}{\mu_D} \sqrt{\frac{2 \epsilon_X}{\mu_D}}\right)^{1/2}\,.
\end{equation}
Here, the reduced mass of $D\olsi D$ pair reads $\mu_D=m_D m_{\olsi D}/(m_D+m_{\olsi D})$, and the binding energy is defined as $\epsilon_X=m_D+m_{\olsi D}-M_{X_0}$ with $M_{X_0}$ the $X_0$ mass. The relativistic coupling is $\chi={\cal N}\chi_{\mathrm{nr}}$ with a normalization factor ${\cal N} =\sqrt{M_{X_0}m_Dm_{\olsi D}}$.

The interaction between the light vector mesons and charmed mesons is described by an effective Lagrangian constructed within the framework of HQSS and hidden local symmetry. Its explicit form reads~\cite{Wise:1992hn,Yan:1992gz,Burdman:1992gh,Casalbuoni:1996pg,Cheng:2004ru}
\begin{align} \label{eq:LagV}
    \mathcal{L}_V &= -\ii g_{DDV} D^\dagger_i \pararrowk{\mu} D^j (V^\dagger_\mu)^i_j \nonumber\\
    &- 2f_{D^*DV} \varepsilon_{\mu\nu\alpha\beta} (\partial^\mu V^{\nu\dagger})^i_j (D^\dagger_i \pararrowk{\alpha} D^{*\beta j} \nonumber\\
    & - D^{*\beta\dagger}_i \pararrowk{\alpha} D^j ) + \ii g_{D^*D^*V} D^{*\nu\dagger}_i \pararrowk{\mu} D^{*j}_\nu (V^\dagger_\mu)^i_j \nonumber\\
    &+ \ii 4f_{D^*D^*V} D^{*\dagger}_{i\mu} (\partial^\mu V^{\nu\dagger} - \partial^\nu V^{\mu\dagger})^i_j D^{*j}_\nu\,,
\end{align}
where $D^{(*)} = (D^{(*)0},D^{(*)+},D^{(*)+}_s)$ stands for the charmed-meson triplet. The light vector-meson matrix $V$ is
\begin{align} \label{eq:matrixV}
V=
\begin{pmatrix}
\frac{\rho^0}{\sqrt{2}}+\frac{\omega}{\sqrt{2}} & \rho^+ & K^{*+}\\
\rho^- & -\frac{\rho^0}{\sqrt{2}}+\frac{\omega}{\sqrt{2}} & K^{*0}\\[6pt]
K^{*-} & \olsi{K}{}^{*0} & \phi
\end{pmatrix}\,.
\end{align}
Based on the HQSS, the coupling constants are linked to each other by the following relation~\cite{Casalbuoni:1996pg,Cheng:2004ru}:
\begin{align}
   g_{DDV} &= g_{D^*D^*V} = \frac{\beta g_V}{\sqrt{2}}\,,\nonumber\\
   f_{D^*DV} &= \frac{f_{D^*D^*V}}{m_{D^*}} = \frac{\lambda g_V}{\sqrt{2}},
\end{align}
where $\beta=0.9$, $\lambda=0.56~\mathrm{GeV}^{-1}$~\cite{Isola:2003fh}, $g_V=m_\rho/f_\pi$ with $m_\rho=0.775~\mathrm{GeV}$ and $f_\pi=0.132~\mathrm{GeV}$~\cite{Casalbuoni:1996pg}.

The electromagnetic interactions between the charmed mesons and the photon comprise both the magnetic and electric contributions. The magnetic coupling between vector and pseudoscalar charmed mesons is given by~\cite{Dong:2009uf,Chen:2010re}
\begin{align} \label{eq:LagDstarDgamma}
    \mathcal{L}_{D^*D\gamma} &= (\frac{\ee}{4} g_{D^{*+}D^+\gamma} \varepsilon^{\mu\nu\alpha\beta} F_{\mu\nu} D^{*+}_{\alpha\beta} D^- \nonumber\\
    &+ \frac{\ee}{4} g_{D^{*0}D^0\gamma} \varepsilon^{\mu\nu\alpha\beta} F_{\mu\nu} D^{*0}_{\alpha\beta} \olsi{D}{}^0) + \mathrm{H.c.}\,,
\end{align}
where $F_{\mu\nu}=\partial_\mu A_\nu-\partial_\nu A_\mu$ is the electromagnetic field strength tensor, $e$ is the elementary electric charge, and $D^*_{\alpha\beta}=\partial_\alpha D^*_\beta-\partial_\beta D^*_\alpha$. 
The coupling constants are $g_{D^{*+}D^+\gamma} = -0.5~\mathrm{GeV}^{-1}$ and $g_{D^{*0}D^0\gamma} = 2.0~\mathrm{GeV}^{-1}$, which are extracted from the experimental widths of $D^{*+}\to D^+\gamma$ and $D^{*0}\to D^0\gamma$~\cite{Chen:2010re,Chen:2015igx}.
The electric couplings are obtained via the gauge transformation $\partial_{\mu}\to\partial_{\mu}+ieA_{\mu}$, leading to the $D^{(*)}D^{(*)}\gamma$ interactions~\cite{Dong:2009uf,Chen:2010re}
\begin{align}
    \mathcal{L}_{DD\gamma} &= \ii \ee A_\mu D^- \pararrowk{\mu} D^+\,,\label{eq:LagDDgamma}\\
    \mathcal{L}_{D^*D^*\gamma} &= -\ii \ee A_\mu (g^{\alpha\beta} D^{*-}_\alpha \pararrowk{\mu} D^{*+}_\beta - g^{\mu\beta}
    D^{*-}_\alpha \partial^\alpha D^{*+}_\beta  \nonumber\\ &+ g^{\mu\alpha} \partial^\beta D^{*-}_\alpha D^{*+}_\beta)\, .\label{eq:LagDstarDstargamma}
\end{align}
After gauging the interaction of Eq.~\eqref{eq:LagV}, we obtain the Lagrangian for the $D\olsi D\gamma V$ coupling
\begin{align} \label{eq:LagDDgammaV}
    \mathcal{L}_{D\olsi D\gamma V} = g_{D D\gamma V} D^\dagger_i A^\mu D^j (V^\dagger_\mu)^i_j\,.
\end{align}
where the coupling constant $g_{D D\gamma V}=2 \ii \ee_D g_{D DV}$ with $\ee_D$ the electric charge of the charmed meson.

\begin{figure*}[tbp]
    \subfigure[ ]{\includegraphics[width=0.28\linewidth]{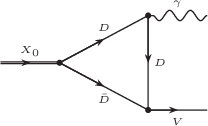}} {\hglue 0.4cm}
    \subfigure[ ]{\includegraphics[width=0.28\linewidth]{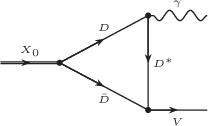}} {\hglue 0.4cm}
    \subfigure[ ]{\includegraphics[width=0.28\linewidth]{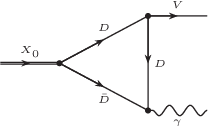}}\\[5pt]
    \subfigure[ ]{\includegraphics[width=0.28\linewidth]{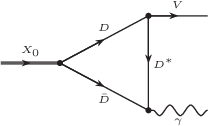}} {\hglue 0.4cm}
    \subfigure[ ]{\raisebox{0.2cm}{\includegraphics[width=0.28\linewidth]{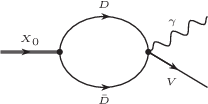}}} {\hglue 0.4cm}
    \subfigure[ ]{\raisebox{0.6cm}{\includegraphics[width=0.16\linewidth]{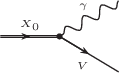}}}
    \caption{Feynman diagrams for the radiative decay $X_0\to\gamma V$. Diagrams (b) and (d) receive loop contributions from both charged $(D^{(*)+}/D^{(*)-})$ and neutral $(D^{(*)0}/\olsi D{}^{(*)0})$ charmed mesons, whereas diagrams (a), (c), and (e) involve only charged charmed-meson loops.}
    \label{fig:feyndiagram}
\end{figure*}

\subsection{Radiative transition amplitudes}

As is well established, the production and decay of the hadronic molecules are described within the hadronic loop mechanism~\cite{Faessler:2007gv,Guo:2013zbw,Li:2014gxa,Chen:2015igx,Shi:2023mer,Wu:2023rrp,Liu:2024ogo,Cai:2024glz}. In this work, we adopt this mechanism to investigate the radiative decay $X_0\to\gamma V$. From the aforementioned effective Lagrangians, we construct the corresponding Feynman diagrams for this decay process, as shown in Fig.~\ref{fig:feyndiagram}. Notably, the diagram (e) is essential to ensure gauge invariance of the decay amplitude~\cite{Faessler:2007gv,Chen:2013cpa,Guo:2014taa,Chen:2015igx,Shi:2023mer,Duan:2024zuo,Gao:2024qth,Liu:2024tgq}. Accordingly, the total amplitude $\mathcal{M}^{\mathrm{tot}}$ for the decay $X_0(p)\to\gamma(p_3) V(p_4)$ encodes the contributions of all the diagrams (a)-(e) in Fig.~\ref{fig:feyndiagram},
\begin{align} \label{eq:amplitude}
    \mathcal{M}^{\mathrm{tot}} = \mathcal{M}^{(a)}+\mathcal{M}^{(b)}+\mathcal{M}^{(c)}+\mathcal{M}^{(d)}+\mathcal{M}^{(e)}\,.
\end{align}

For each topology, $i=a,b,c,d,e$, the corresponding contribution takes the form
\begin{align}
\mathcal{M}^{(i)} = \chi \epsilon^{*\mu}(\gamma) \epsilon^{*\nu}(V) \mathcal{M}^{(i)}_{\mu\nu}\,,
\end{align}
where $\epsilon^{*\mu}(\gamma)$ and $\epsilon^{*\nu}(V)$ denote the polarization vectors of the photon and vector meson, respectively, with $V=\rho^0,\omega$. Explicitly, the tensor structures associated with the individual topological diagrams are given by
\begin{align} \label{eq:amp_a}
    \mathcal{M}^{(a)}_{\mu\nu} &= \ii^3 \int\frac{d^4 q}{(2\pi)^4} \{\ii\ee [-\ii (p_3+q)_\mu - \ii q_\mu]\} \nonumber\\
    &\times \{\ii g_{DDV} [\ii q_\nu - \ii (p_4-q)_\nu]\} \nonumber\\
    &\times S(p_3+q) S(p_4-q) S(q) \mathcal{F}(q^2)\,,
\end{align}

\begin{align} \label{eq:amp_b}
    \mathcal{M}^{(b)}_{\mu\nu} &= \ii^3 \int\frac{d^4 q}{(2\pi)^4} \{\ee g_{D^*D\gamma} \varepsilon_{\sigma\mu\lambda\xi} (\ii p^\sigma_3) (\ii q^\lambda)\} \nonumber\\
    &\times \{2f_{D^*DV} \varepsilon_{\alpha\nu\beta\rho} (\ii p_4^\alpha) [\ii q^\beta - \ii (p_4-q)^\beta]\} \nonumber\\
    &\times S(p_3+q) S(p_4-q) S^{\xi\rho}(q) \mathcal{F}(q^2)\,,
\end{align}

\begin{align} \label{eq:amp_c}
    \mathcal{M}^{(c)}_{\mu\nu} &= \ii^3 \int\frac{d^4 q}{(2\pi)^4} \{\ii g_{DDV} [\ii (p_4+q)_\nu + \ii q_\nu]\} \nonumber\\
    &\times \{\ii \ee (-\ii q_\mu + \ii (p_3-q)_\mu)\} \nonumber\\
    &\times S(p_4+q) S(p_3-q) S(q) \mathcal{F}(q^2)\,,
\end{align}

\begin{align} \label{eq:amp_d}
    \mathcal{M}^{(d)}_{\mu\nu} &= \ii^3 \int\frac{d^4 q}{(2\pi)^4} \{2f_{D^*DV} \varepsilon_{\alpha\nu\beta\rho} (\ii p_4^\alpha) [-\ii (p_4 \nonumber\\
    &+ q)^\beta - \ii q^\beta)\} \{\ee g_{D^*D\gamma} \varepsilon_{\sigma\mu\lambda\xi} (\ii p^\sigma_3) (-\ii q^\lambda)\} \nonumber\\
    &\times S(p_4+q) S(p_3-q) S^{\rho\xi}(q) \mathcal{F}(q^2)\,,
\end{align}

\begin{align} \label{eq:amp_e}
    \mathcal{M}^{(e)}_{\mu\nu} &= \ii^2 \int\frac{d^4 q}{(2\pi)^4} \{-\ii g_{DD\gamma V} g_{\mu\nu}\} \nonumber\\
    &\times S(p_3+p_4-q) S(q) \mathcal{F}_{\mathrm{Con}}(q^2)\,,
\end{align}
where the propagators of $D$ and $D^*$ mesons are
\begin{align}
    S(p) &= \frac{1}{p^2-m_D^2 + \ii \epsilon}\,, \label{eq:prop_D}\\
    S^{\mu\nu}(k) &= \frac{-g^{\mu\nu} + k^\mu k^\nu/m_{D^*}^2}{k^2 - m_{D^*}^2 + \ii \epsilon}\,. \label{eq:prop_Dstar}
\end{align}
Here, we adopt the dipole form factor to regularize the ultraviolet divergence of the loop functions $\mathcal{M}_{\mu\nu}$,
\begin{equation} \label{eq:formfactor}
   \mathcal{F}(q^2) = \left( \frac{m^2-\Lambda^2}{q^2-\Lambda^2} \right)^2,
\end{equation}
where $m$ and $q$ represent the mass and four-momentum of the intermediate exchanged mesons, respectively. The cutoff $\Lambda$ is parametrized as $m+\alpha\Lambda_{\mathrm{QCD}}$, where $\Lambda_{\mathrm{QCD}}=220~\mathrm{MeV}$ and $\alpha$ is a dimensionless parameter. This form factor effectively accounts for the internal structure of interaction vertices and the off-shell effects of exchanged mesons. We also introduce an additional form factor $\mathcal{F}_{\mathrm{Con}}(q^2)$~\cite{Chen:2013cpa,Duan:2024zuo}, which is required to enable the gauge invariance of the total amplitude.

We now explicitly verify the Ward identity for the decay amplitude. After reducing the loop integrals, the Lorentz structures of the amplitudes corresponding to diagrams (a)-(e) in Fig.~\ref{fig:feyndiagram} can be decomposed as
\begin{align}
    \mathcal{M}^{(a)}_{\mu\nu} &= \mathcal{M}^{(c)}_{\mu\nu} = (A_1 g_{\mu\nu} p_3 \cdot p_4 + A_2 p_{3\nu} p_{4\mu}) \, ,\nonumber\\
    \mathcal{M}^{(b)}_{\mu\nu} &= \mathcal{M}^{(d)}_{\mu\nu} = B (g_{\mu\nu} p_3 \cdot p_4 - p_{3\nu} p_{4\mu})\,,\nonumber\\
\mathcal{M}^{(e)}_{\mu\nu} &= C g_{\mu\nu} p_3 \cdot p_4,
\end{align}
where the coefficients $A_1$, $A_2$, and $B$ are obtained from the triangle diagrams, while $C$ denotes the contribution from the two-point loop diagram (e).

It is straightforward to see that the contributions from the photon magnetic coupling, $\mathcal{M}^{(b)}_{\mu\nu}$ and $\mathcal{M}^{(d)}_{\mu\nu}$, are individually gauge invariant, $i.e.$, $p_3^\mu \mathcal{M}^{(b)}_{\mu\nu}=p_3^\mu \mathcal{M}^{(d)}_{\mu\nu}=0$. Consequently, contracting the total amplitude with the photon momentum $p_3^\mu$ gives
\begin{align}
p_3^\mu \mathcal{M}^{\mathrm{tot}}_{\mu\nu}=p_{3\nu} p_3 \cdot p_4\left(2A_1+2A_2+C\right).
\end{align}
The Ward identity requires
\begin{equation}
    C = -2(A_1 + A_2)\,.
\end{equation}
We have constructed the form factor $\mathcal{F}_{\mathrm{Con}}(q^2)$ to ensure the total amplitude satisfies the above relation.

The total amplitude $\mathcal{M}^{\mathrm{tot}}$ in Eq.~\eqref{eq:amplitude} is a UV divergent integral. To ensure the total amplitude is well-defined, we introduce a $X_0\gamma V$ counterterm amplitude for the decay $X_0\to \gamma V$, depicted in Fig.~\ref{fig:feyndiagram}(f), 
\begin{align}
{\cal M}^{\text{Con}}=i\left(\lambda_0 g_{\mu\nu} + \lambda_1 p_{3\nu} p_{4\mu}\right)\epsilon^{*\mu}(\gamma) \epsilon^{*\nu}(V),
\label{Eq:counterterm}
\end{align}
where the amplitude ${\cal M}^{\text{Con}}$ is required to satisfy the Ward identity. Here $\lambda_0$ and $\lambda_1$ are unknown constants, which are used to absorb the UV divergence from the loops in Eq.~\eqref{eq:amplitude}.

The final decay width for the process $X_0\to\gamma V$ can then be written as
\begin{equation}
    \Gamma(X_0\to\gamma V) = \frac{1}{8\pi} \frac{p}{M_{X_0}^2} \sum_{\text{polarizations}}|\mathcal{M}^{\mathrm{tot}}|^2\,,
\end{equation}
where $p=|\mathbf{p}|$ denotes the momentum of the final states in the center-of-mass frame.

\section{Numerical results} \label{sec:result}

In this section, we calculate the partial widths for the radiative decays of the $X_0\to\gamma\rho^0$ and $X_0\to\gamma\omega$. The masses of charmed and light mesons are taken from the PDG~\cite{ParticleDataGroup:2024cfk}, while the mass of $X_0$ remains undetermined. Under the assumption of a shallow bound state, we adopt the criterion $E_B<1/(2\mu R_{\mathrm{conf}}^2)$ with $R_{\mathrm{conf}}<1$~fm~\cite{Guo:2017jvc} and then constrain the binding energy in the range of $\epsilon_X \sim 0$--$20~\mathrm{MeV}$. In addition, we examine the dependence of the decay widths on the cutoff parameter $\alpha$, which is varied within the range $\alpha=0.6$--$1.2$, following Refs.~\cite{Chen:2010re,Wu:2021udi,Zheng:2024eia}.

\begin{figure}
	\centering
	\includegraphics[width=0.98\linewidth]{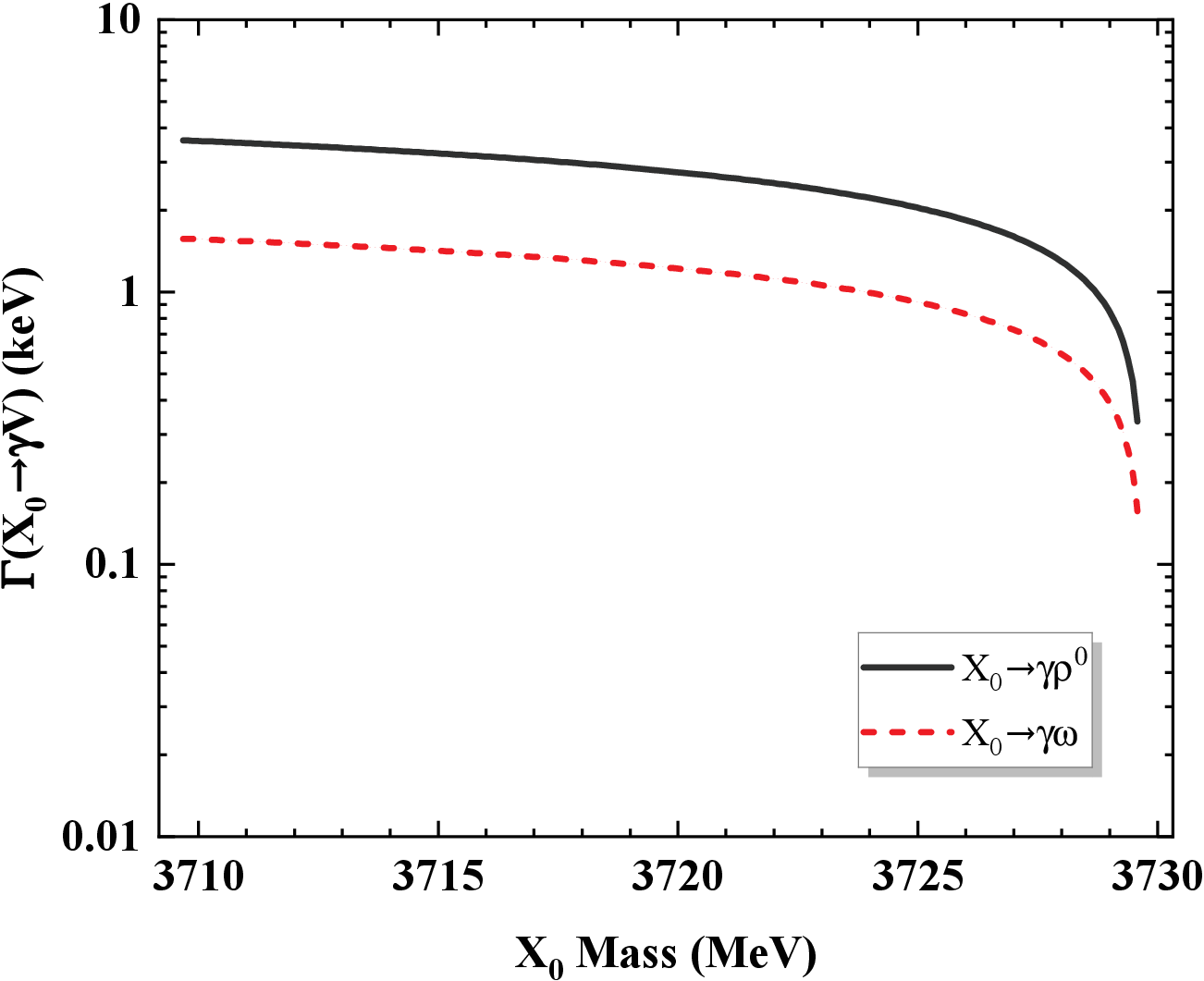}
	\caption{Partial widths for the processes $X_0\to\gamma\rho^0$ and $X_0\to\gamma\omega$ as functions of the $X_0$ mass with the model parameter fixed at $\alpha=0.9$.}
	\label{fig:widthmass}
\end{figure}

We investigate the dependence of the partial widths of $X_0\to\gamma\rho^0$ and $X_0\to\gamma\omega$ on the $X_0$ mass. Fixing the cutoff parameter at the intermediate value $\alpha=0.9$, we present the resulting partial widths in Fig.~\ref{fig:widthmass}. As the $X_0$ mass increases, the partial widths of both radiative decay channels decrease. This tendency is caused by the effective coupling $\chi_{\text{nr}}$, which is sensitive to $M_{X_0}$.

\begin{figure}
	\centering
	\includegraphics[width=0.98\linewidth]{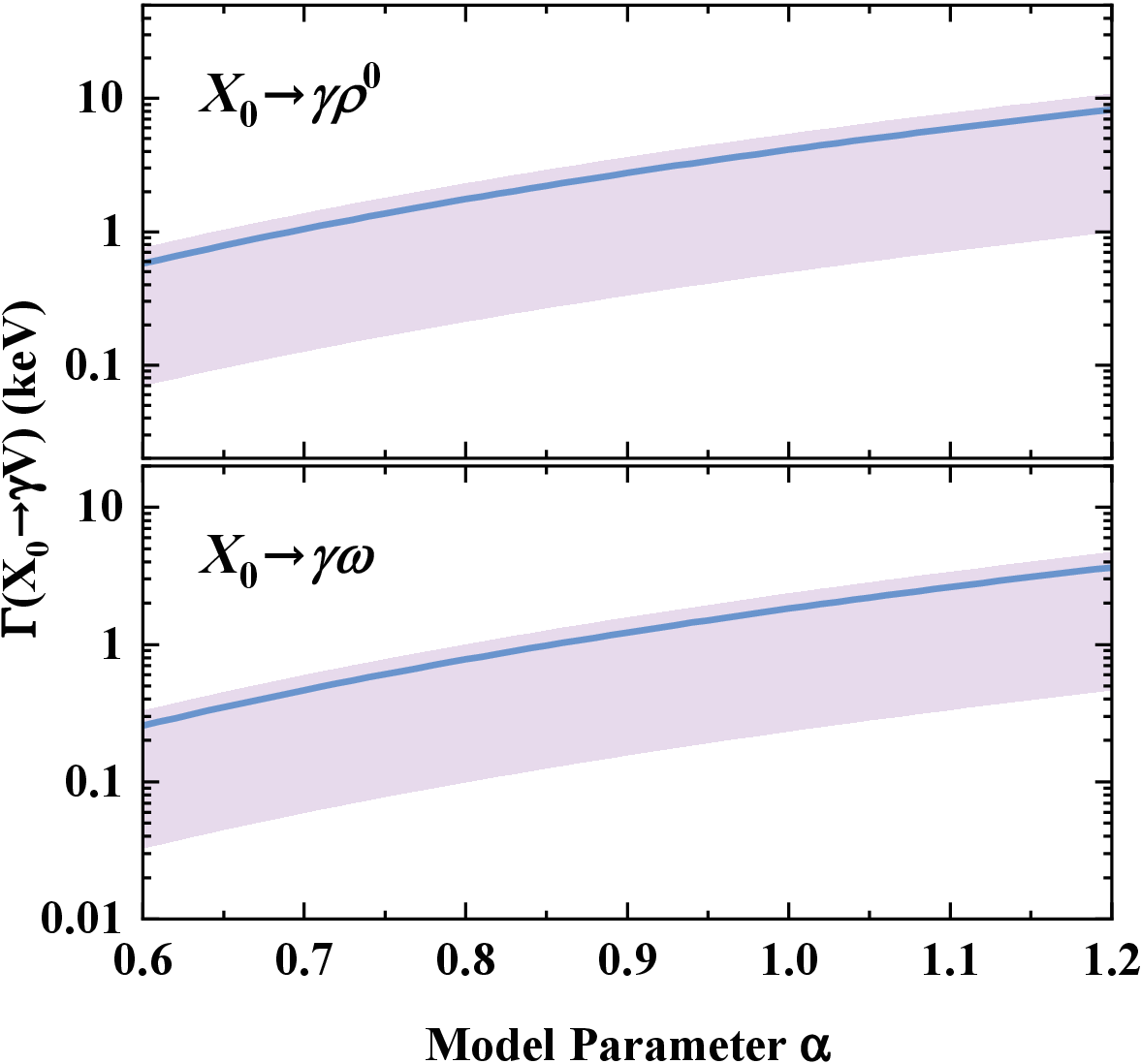}
	\caption{Partial widths of $X_0\to\gamma V$ as a function of the cutoff parameter $\alpha$. The blue solid curve corresponds to the result obtained with $M_{X_0}=3720~\mathrm{MeV}$, and the purple band represents the uncertainty from $M_{X_0}=3709.68$--$3729.58~\mathrm{MeV}$.}
	\label{fig:widthalpha}
\end{figure}

The partial widths also depend on the cutoff parameter. As shown in Fig.~\ref{fig:widthalpha}, the partial widths of $X_0\to\gamma V$ exhibit a significant sensitivity to $\alpha$. For illustration, taking $M_{X_0}=3720~\mathrm{MeV}$~\cite{Wei:2022jgc}, the widths increase by approximately an order of magnitude as $\alpha$ is varied from $0.6$ to $1.2$:
\begin{align} \label{eq:width}
    \Gamma(X_0\to\gamma\rho^0) &= 0.58-8.23~\mathrm{keV}\,,\nonumber\\
    \Gamma(X_0\to\gamma\omega) &= 0.26-3.64~\mathrm{keV}\,.
\end{align}
The pronounced cutoff dependence indicates that the loop contributions retain significant regulator sensitivity. Such dependence should be absorbed by the $X_0\gamma V$ counterterms, depicted in Fig.~\ref{fig:feyndiagram}(f), to ensure that the physical decay widths are cutoff-independent. Since the low-energy constants, $\lambda_0$ and $\lambda_1$ in Eq.~\eqref{Eq:counterterm}, are presently unknown, the absolute values of the radiative widths cannot be predicted in a regulator-independent manner from the loop contributions.

\begin{figure}[htbp]
	\centering
	\includegraphics[width=0.98\linewidth]{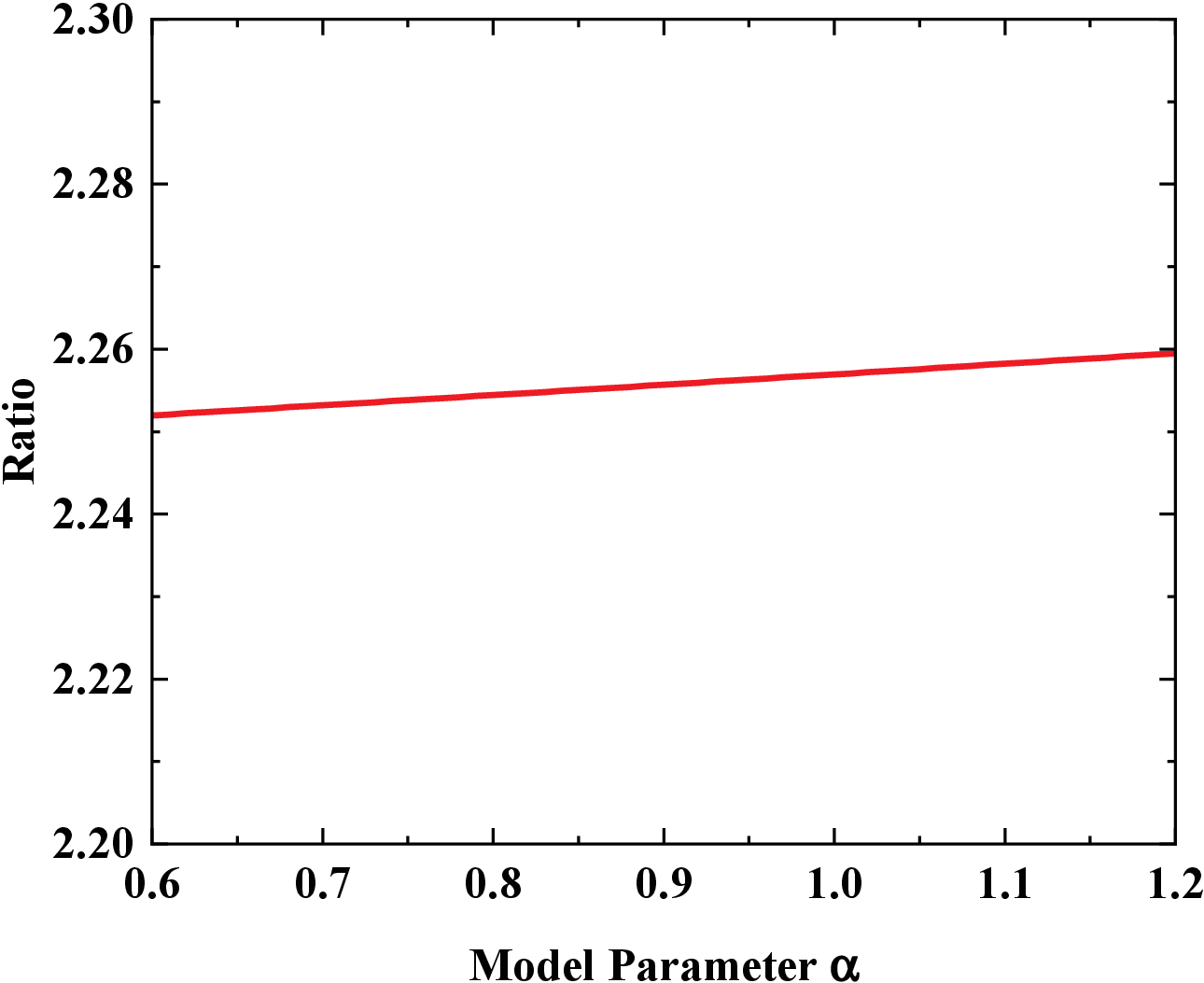}
	\caption{Partial decay width ratio $\Gamma(X_0\to\gamma\rho^0)/\Gamma(X_0\to\gamma\omega)$ as a function of the model parameter $\alpha$.}
	\label{fig:ratio}
\end{figure}

To reduce this model dependence, we therefore focus on the ratio of the two radiative decay widths. In the following, we set the unknown low-energy constants, $\lambda_0$ and $\lambda_1$, to zero and examine the ratio
\begin{align}
R=
\frac{\Gamma(X_0\to\gamma\rho^0)}
{\Gamma(X_0\to\gamma\omega)}.
\end{align}
Although the individual widths retain a sizable dependence on the cutoff parameter, the cancellation of the regulator dependence can occur in their ratio because the two channels are generated by closely related loop mechanisms~\cite{Guo:2014taa,Shi:2023mer}.
The resulting ratio is displayed in Fig.~\ref{fig:ratio}. The obtained ratio remains remarkably stable over the considered range of $\alpha$, with a relative fluctuation of about $0.33\%$. Such weak parametric dependence supports the validity of our theoretical model. This weak cutoff dependence  provides a more reliable observable for radiative decay of $X_0$.

The small cutoff dependence of ratio $R$ indicates that we can choose any value of the ratio in the large $\alpha$ region as our prediction. We therefore take the ratio $R=2.26$ as our theoretical prediction. The large value of $R$ originates from a pronounced interference effect between the charged and neutral charmed-meson loop contributions. These contributions largely cancel in the $X_0\to \gamma \omega$, whereas the interference effect lead to a enhancement in the $X_0\to \gamma \rho^0$. This pronounced difference between the two decay modes highlights the sensitivity of the radiative $X_0$ decays to isospin-breaking channel. Our results demonstrate that these decay modes provide a sensitive probe of the interplay between charged and neutral hadronic components and offer valuable insight into the underlying hadronic-loop dynamics and the molecular structure of $X_0$.

\section{Summary} \label{sec:summary}

In this work, we have investigated the radiative decays of the isoscalar $S$-wave $D\olsi D$ molecular state $X_0$, as predicted within the framework of HQSS. The radiative transitions $X_0\to \gamma V$ with $V=\rho^0,\omega$ proceed through intermediate charmed-meson loop diagrams. In our calculation, we consistently impose electromagnetic gauge invariance to ensure that the total transition amplitudes satisfy the Ward identity. For a fixed $M_{X_0}=3720~\mathrm{MeV}$, we obtained the partial widths of $0.58-8.23~\mathrm{keV}$ for $X_0\to\gamma\rho^0$ and $0.26-3.64~\mathrm{keV}$ for $X_0\to\gamma\omega$, as the model parameter $\alpha$ varies from $0.6$ to $1.2$. We also find that both partial widths decrease as the $X_0$ mass increases within the shallow bound state range. Remarkably, while the individual decay widths exhibit a sizable dependence on the model parameter, their ratio remains essentially insensitive to variations of $\alpha$. We therefore predict that the ratio of $X_0\to \gamma \rho^0$ and $X_0\to \gamma \omega^0$ is approximately $2.26$, which is nearly independent on the cutoff. This robust ratio provides a useful observable for probing the molecular nature of $X_0$ and offers a relatively model-independent prediction for future experimental studies.

\begin{acknowledgments}
\label{sec:acknowledgements}
This work is supported by the National Natural Science Foundation of China under Grant No. 12475081; by the Natural Science Foundation of Shandong Province under Grant No. ZR2025MS04; and by Taishan Scholar Project of Shandong Province under Grant No. tsqn202607074. This work is also supported by the Grants {\small PID2023-147458NB-C21} and {\small CEX2023-001292-S} funded by {\small MICIU/AEI/10.13039/501100011033} and by ERDF/EU, as well as of the Grant {\small CIPROM/2023/59} funded by Generalitat Valenciana {\small 10.13039/501100003359}.
\end{acknowledgments}

\bibliography{references.bib}
\end{document}